\documentclass{ws-p8-50x6-00}

\newcommand{\nuc}[2]{\mbox{\textsuperscript{#1}#2}}
\newcommand{\g}{\ensuremath{\gamma}}

\newcommand{\keV}{\mbox{ke\/V}}

\newcommand{\us}{\mbox{$\mu$s}}

\begin{document}

\title{XENON: a 1 tonne Liquid Xenon Experiment for a Sensitive Dark Matter
Search} 

\author{E.~Aprile, E.A.~Baltz, A.~Curioni, K-L.~Giboni, C.J.~Hailey, L.~Hui,
M.~Kobayashi, K.~Ni }

\address{Columbia University \\ 
E-mail: age@astro.columbia.edu}

\author{W.W.~Craig}

\address{Lawrence Livermore National Laboratory}

\author{R.J.~Gaitskell}

\address{Brown University}

\author{U.~Oberlack}

\address{Rice University}  

\author{T.~Shutt}

\address{Princeton University}

\maketitle


\abstracts{XENON is a novel liquid xenon experiment concept for a sensitive dark
matter search using a 1-tonne active target, distributed in an array of ten independent
time projection chambers.  
The design relies on the simultaneous detection of ionization and scintillation
signals in liquid xenon, with the goal of extracting as much information as
possible on an event-by-event basis, while maintaining  most of the target
active. XENON is expected to have effective and redundant background
identification and discrimination power, higher than 99.5\%, and to achieve a
very low threshold, on the order of 4~keV visible recoil energy. 
Based on this expectation and the 1-tonne mass of active xenon, we project a
sensitivity of 0.0001~events/kg/day, after 3~yr
operation in an appropriate underground location. The XENON experiment has been
recently proposed  to the National Science Foundation (NSF) for an initial
development phase leading to the development of the 100 kg unit module.} 

\section{Introduction}

Substantial astronomical evidence shows that at least 90\% of the mass in the
universe is dark, and that most of it is non-baryonic in nature (see e.g.
reviews \cite{trimble87,primack88,tremaine92,jungman96}).  Dark matter plays a
central role in current structure formation theories, and its microscopic
properties have a significant impact on the spatial distribution of mass,
galaxies and clusters. Unraveling the nature of dark matter is therefore of
critical importance. Several lines of arguments indicate that the dark matter
consists of Weakly Interacting Massive Particles (WIMPs), a well-motivated
example of which is the neutralino, the lightest supersymmetric particle. Direct
detection, via elastic scattering of a WIMP on a suitable target, offers the
hope of studying the dark matter properties in detail, and shedding light on
particle physics beyond the Standard Model. 


In spite of the experimental challenges, a number of efforts worldwide are
actively pursuing to directly detect WIMPs with a variety of targets and
approaches. 
One approach is to decrease the radioactive background to extreme low levels,
using a high purity Germanium target and detector, with careful selection of
surrounding materials \cite{klapdor01,HDMS99,GENIUS98}.  
A second approach, followed by the DAMA \cite{DAMA00} and the UKDM NAIAD
\cite{spooner00} groups, has been to use large NaI scintillators with pulse
shape background discrimination. 
The third experimental approach relies on more powerful discrimination methods,
using various schemes to extract as much information as possible from the
target-detector. To this class belong the cryogenic detectors based on the
simultaneous measurement of ionization and phonons in crystals of Ge or Si, as
used by the CDMS experiment \cite{abusaidi00} and the EDELWEISS experiment
\cite{benoit01}, or phonons and scintillation light in CaWO$_4$ crystals as used
by the CRESST experiment~\cite{bravin99}. 
Experiments based on the simultaneous detection of ionization and scintillation
light in liquid xenon (LXe) belong to the same class and offer an equally
promising approach to direct detection of WIMPS in large scale targets. 
Two experiments, ZEPLIN~II\cite{HWang:00:ZEPLINII} and III\cite{ZEPLINAXE},
with 30~kg and 6~kg of Xe mass, respectively, are currently being developed as part
of the UKDM LXe program. 
Scale-up to the 1-tonne level is in the planning or proposal
stage~\cite{ZEPLINAXE,ZEPLINIV}.  

The XENON project, recently proposed to NSF for an initial development phase, is
an alternative concept for a 1-tonne LXe experiment, to be located in the
National Underground Science Laboratory (NUSL), under discussion in the US.
The goal of the XENON experiment is to achieve a factor of 30 higher
sensitivity than that projected for CDMS~II~\cite{CDMS_Soudan} in the US
and other experiment in Europe (e.g. EDELWEISS). This sensitivity increase is
needed to probe the lowest SUSY predictions for the neutralino.
With 1-tonne target mass, a visible energy threshold of 4 keV and a
background discrimination factor much better than 99.5\%, XENON projected
sensitivity is 0.0001~events/kg/day after 3~yr operation.

\section{Liquid Xenon for Dark Matter}

\begin{figure}
\centering
\includegraphics[width=0.7\linewidth]{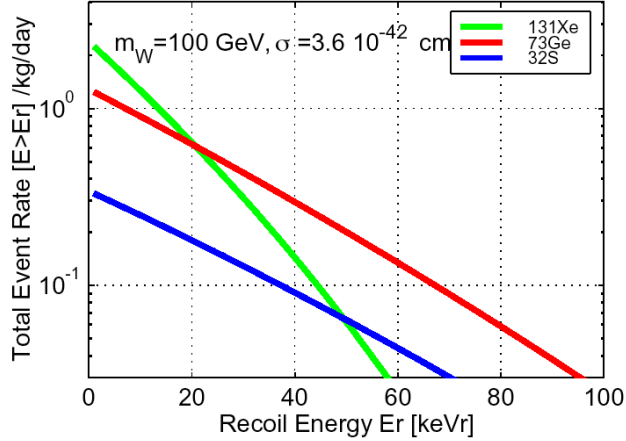}
\caption{Expected event rate, integrated above energy threshold, for Xe, Ge and
S, for a 100~GeV WIMP with cross section $\sigma$~=~3.6 10$^{-42}$~cm$^2$, under
standard assumptions of a dark matter halo.
\label{f:rick}}  
\end{figure}

Liquid xenon is an attractive target for a sensitive WIMP search. Its high
density ($\sim$~3~g/cm$^3$) and high atomic number (Z~=~54, A~=~131) allow for a
compact detector geometry. The high mass of the Xe nucleus is favorable for WIMP
scalar interactions, provided a low recoil energy threshold, as shown in Fig.~\ref{f:rick}.
The expected event rate, integrated above the energy threshold, is calculated for Xe, Ge
and S, assuming a 100~GeV WIMP with a cross-section $\sigma$~=~3.6~10$^{-42}$~cm$^2$
(see~\cite{lewin96} as a standard reference). 
As detector material LXe has excellent ionization and scintillation
properties. With the simultaneous measurement of charge and light and 3D
position resolution, event information can be maximized to achieve effective and
redundant background identification and discrimination power, while maintaining
most of the target active. Xenon, which contains both odd and even isotopes for
coherent and purely spin-dependent WIMP interactions, is available in large
quantities at reasonable cost. Various techniques have demonstrated ultra pure
LXe in which an electron lifetime in excess of 1~ms 
\cite{EAprile:91:lifetime,EAprile:01:small:chamber,Benetti93} allows
the drift of free electrons over 30 cm and longer. The reduction of the krypton
contamination in natural xenon to the required part per billion (ppb) level has
also been verified with a distillation tower and cold traps. 

\section{The XENON Experiment: Design Overview}\label{sec:XENON}

\begin{figure}[t]
\centering
\includegraphics[bb=115 15 693 599,width=0.7\linewidth,clip]{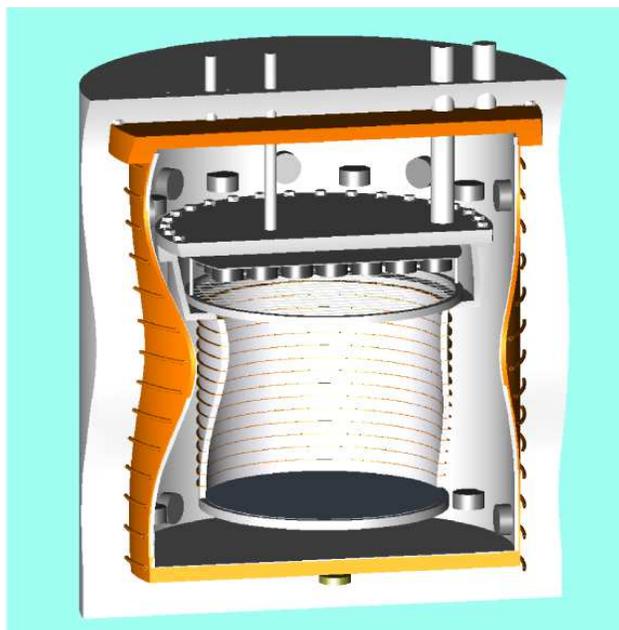}
\caption{The LXeTPC module for XENON: the 100 kg fiducial target is surrounded
by an active LXe shield enclosed in the Cu vessel.\label{f:3D}}
\end{figure}

The XENON design is modular. An array of 10 independent 3D position sensitive
liquid xenon time projection chambers (LXeTPC) makes the 1-tonne scale
experiment. Each TPC contains 100 kg of active Xe mass and is self-shielded
with additional LXe scintillator, as schematically shown in Fig.~\ref{f:3D}.

\begin{figure}[h]
\centering
\includegraphics[width=0.7\linewidth,clip,angle=90]{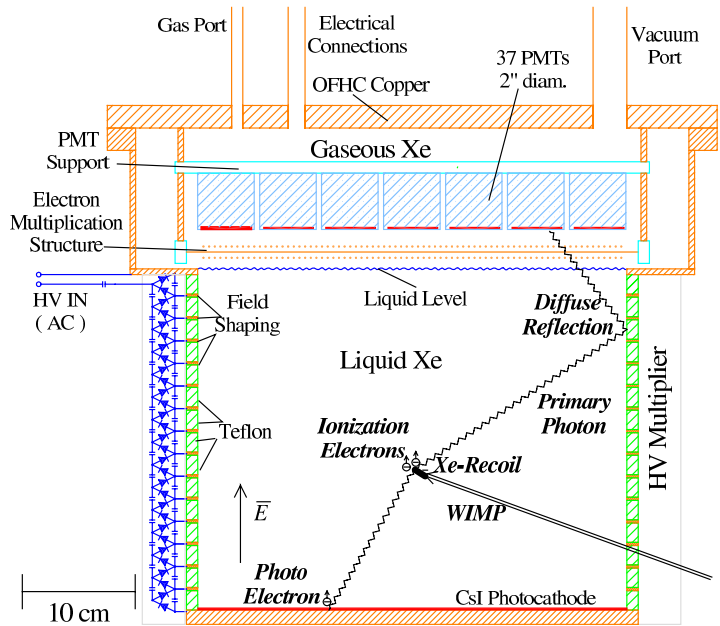}
\caption{The LXeTPC module for XENON: Schematic Design of the 100 Kg detector
and its components.\label{f:Karl:layout}}
\end{figure}

The modular approach is preferred over a single
detector, for several reasons. The most important is a feasibility argument. We
have already built a 30 kg LXeTPC and have used it for several balloon flights
of the Liquid Xenon Gamma-Ray Imaging Telescope (LXeGRIT)
\cite{EAprile:96:LXeGRIT,EAprile:98:electronics,EAprile:2000:cgro99,EAprile:00:spie00:flight99,EAprile:01:daq,UOberlack:00:spie00:CSR,UOberlack:01:trigger}. In
this TPC, with a maximum drift gap of 10~cm, both charge and 
light signals are detected for imaging gamma-rays. The excellent background
discrimination which directly stems from the 3D event localization in the
homogeneous volume has been demonstrated. The experience gained though this
development effort, and the prior years of R\&D on noble liquid detectors
(e.g.\cite{EAprile:90:ITNS,EAprile:91:lifetime,EAprile:91:performance,EAprile:92:2DTPC,EAprile:93:monte_carlo}), 
give us the confidence that a 100~kg TPC optimized for dark matter detection can be built 
successfully.  With an array, a failure of one module
would not halt the entire experiment. Operational 
efficiency is clearly higher than with a monolithic detector of 1-tonne.

As shown in details in Fig.~\ref{f:Karl:layout}, the LXeTPC structure containing
the active Xe target is formed by a sandwich of Teflon spacers as UV diffuse
reflector and copper rings for electric field shaping. 
The structure is closed at the bottom by a copper plate. The inside of this
plate is coated with CsI as photocathode to convert Xe scintillation photons
into free electric charges. The structure forms a 30~cm high cylinder with 38~cm
inner diameter, holding about 100~kg of ultra pure liquid xenon. 

On the top, the structure is hermetically sealed to a cylindrical copper vessel
of larger diameter, housing the PMTs and the wire structure for the proportional
scintillation process in the gas phase.

The LXeTPC structure is enclosed in a copper vessel containing the liquid xenon
for active shielding. With both detectors at the same temperature and similar
pressure, the amount of material for the inner detector walls is minimized. The
scintillation light from the shield section is viewed by two rings of 16 PMTs each.

\subsection{The XENON LXeTPC: Principle of Operation}\label{subsec:POO}

Referring to Fig.~\ref{f:Karl:layout}, which shows the TPC
design in more details, we limit the discussion to the relevant processes in LXe which are used
to detect and discriminate a WIMP from background.   

The elastic scattering of a WIMP within the active target results in a low
energy Xe recoil. The moving recoil produces both ionization electrons and fast
UV scintillation photons at 178~nm, from the de-excitation to the ground state
of excited diatomic Xe molecules (Xe$^{\ast}_2$).   
  
The number of UV photons associated with direct Xe
excitation by a nuclear recoil is only a fraction of that emitted by an
electron or gamma-ray with the same kinetic energy. Recent measurements of this
`quenching factor' for Xe scintillation range from about 22\% \cite{Arneodo00}
to 45\% \cite{Bernabei01}. The number  of free electrons 
liberated by a nuclear recoil is also very small, even if a strong electric
field is applied across the liquid \cite{EAprile:90:ITNS,EAprile:91:alpha},
because the bulk of the ionization electrons recombine within picoseconds. 

Thus, under a high electric field, a nuclear recoil will yield a very small
charge signal and a much larger light signal, compared to an electron
recoil of the same energy. The distinct charge/light
ratio is the basis for nuclear recoil discrimination in a LXe detector. To
detect the small charge signals involved, the process of electroluminescence
\cite{Bolozdynya:99:lumi} is typically used. The free ionization electrons are
extracted from the liquid to the gas phase where in the strong electric field around thin
wires they induce proportional scintillation light. The number of photons
generated by one drifting electron is sufficiently large to be detected by PMTs.
 Results obtained with small size dual
phase LXe prototypes have demonstrated the power of the method
\cite{HWang:98} and motivate the ongoing efforts to use LXe for a direct WIMP
search with high sensitivity.  
The challenge ahead lies in detecting both charge and light signals with high
efficiency down to the lowest possible energy threshold. 
Furthermore, this has to be realized in a detector of sizeable scale and with
the highest ratio of active/passive LXe. 

These considerations have guided our design of the XENON TPC. In the baseline
concept the primary UV photons are detected by an array of PMTs 
placed above the liquid-gas interface. To increase the solid angle and thus
detection efficiency, a CsI photocathode deposited on the bottom plate is used to
convert downwards heading photons into photoelectrons. The efficient
extraction of photoelectrons from CsI in liquid rare gases was originally
demonstrated by the Columbia group \cite{EAprile:94:CsI:perf}. 

\begin{figure}
\centering
\includegraphics[width=0.7\linewidth]{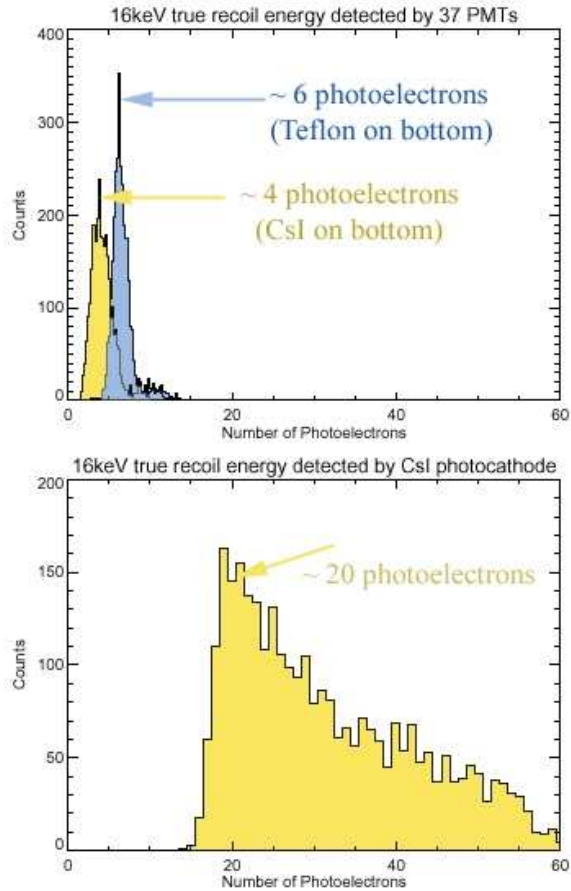}
\caption{Number of photoelectrons for a 16~keV nuclear recoil. {\it Top:} as
seen by the PMTs, {\it bottom:} extracted from the CsI
photocathode. \label{f:PE}}  
\end{figure}

To further increase the primary light collection efficiency, which ultimately
determines the detector energy threshold, the TPC walls are made of
Teflon which has about 90\% diffuse reflectivity at 178~nm \cite{Yamashita00}.
At the liquid-gas interface, the charges are extracted into the gas phase and are
detected via the proportional scintillation signal induced around thin wires
(Fig.~\ref{f:PE} shows the expected number of photoelectrons for a 16~keV
``true'' nuclear recoil. The CsI readout contributes the largest fraction).  

\begin{figure}[htb]
\centering
\includegraphics[height=0.7\linewidth,angle=90]{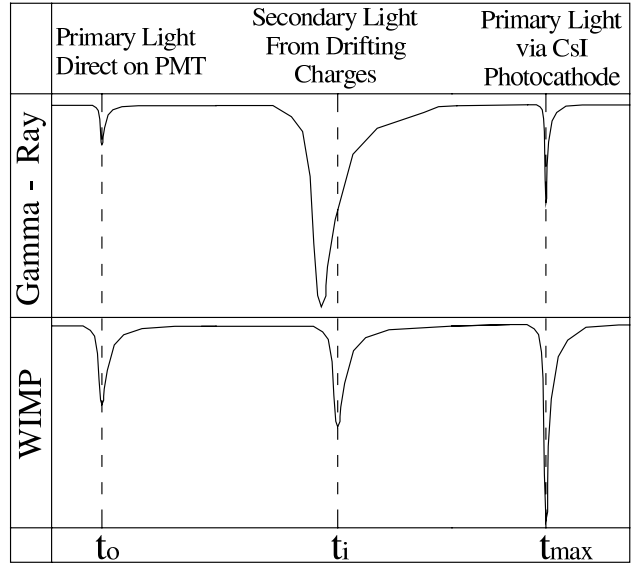}
\label{f:Karl:timepro}
\caption{Shaped light profiles from an electron recoil (top) 
and nuclear recoil (bottom) in the XENON LXeTPC.}
\end{figure}

Most events will provide three signals, with a timing scheme  shown schematically in
Fig.~\ref{f:Karl:timepro}. The first is the prompt scintillation signal detected
directly by the PMTs. The last is the proportional scintillation signal from
the CsI photoelectrons drifting the entire 30 cm liquid gap. These two signals are
separated by exactly 150~\us, i.e. the maximum drift time. The proportional scintillation signal from the
drift of ionization electrons can occur anywhere in between these two. The
difference in arrival time between the primary scintillation pulse and the
proportional pulse from the electron drift measures the interaction depth
(Z-coordinate) of the event. Since electron diffusion in LXe is small, the
proportional scintillation pulse is produced in a small spot with the same X-Y
coordinates as the interaction site. The photons in the proportional
scintillation will spread over several PMTs in the vicinity of this spot. By a
center of gravity method, the X-Y position can be reconstructed to about 1~cm
precision. The X-Y information, along with the absolute Z, gives a
3D localization which will permit further background discrimination via fiducial
volume cuts.  For events where the first
pulse is below detection threshold,  the third pulse will always be present since
the CsI has high quantum efficiency and the signal is amplified by proportional
scintillation. The Z-coordinate can still be inferred from the relative drift
time difference, since the third pulse contains the same information as the
first. In such events with incomplete signatures, the sum signal from the wire structure will be
present only if a proportional scintillation pulse is detected. 
Based on the redundant information in our design and  the 3D
position sensitivity,  we expect to achieve a background rejection efficiency
better than 99.5\% and energy threshold as low as few~keV. 

\section{Background Considerations}\label{sec:BKGD}

To maximize the benefit of the various background rejection factors available
for the XENON experiment, the absolute count rate itself must be minimized. This
is a challenge for any WIMP experiment, as the possible sources of background
are numerous and of different origin. Here, we limit the discussion to the
dominant sources specific to a LXe experiment. Natural Xe has no long-lived
isotopes. However, radioactive impurities in the gas, most notably
\nuc{85}{Kr}, must be reduced by a large amount.

\subsection{Gamma and Beta Induced Background}\label{subsec:GBIB}

\noindent {\em \nuc{85}{Kr} and Radon} --
Commercially available, research grade xenon gas, typically has a Krypton
fraction of 5-10~ppm. Kr has two long-lived radioisotopes, \nuc{81}{Kr} and
\nuc{85}{Kr}. Due to its longer lifetime and smaller abundance, the background
from \nuc{81}{Kr} is weaker than that from \nuc{85}{Kr} by a factor of several
$10^5$. Most of the \nuc{85}{Kr} found in the environment today was released by
nuclear-fuel reprocessing plants, as a fission product of \nuc{235}{U} and
\nuc{239}{Pu}. Its abundance is $\sim 2\times10^{-11}$~\cite{Collon97}.  
The \nuc{85}{Kr} $\beta$-decays with an
endpoint energy of 678~keV. With a half-life of 10.8~yr, the residual signal in
the XENON detector energy band is $\sim$ 20 cts/kg/d/keV.  Reducing this to
$\sim 1\times10^{-5}$~cts/kg/d/keV, with the assumed rejection power of the
detector, requires a concentration of Kr in Xe of 1~ppb. This can be achieved by
distillation and cold traps. The \nuc{85}{Kr} level will be monitored in
a similar way as in the BOREXINO experiment \cite{Borexino}, namely through the
delayed gamma-rays of a rare decay \cite{Alimonti98}.  Radon, continuously
produced by the decay chain of uranium contained in detector materials, will
also be effectively removed with gas re-circulation and cold traps.

\noindent {\em \nuc{136}{Xe} double beta decay} --
Assuming a lifetime of $8 \times 10^{21}$ years and Q=2.48~MeV we can use the
double beta spectrum and directly calculate the fraction of events in our energy
band of interest. 
The resulting count rate is $1\times10^{-6}$~cts/kg/d/keV, which is small
compared to other sources of background. And this is the count rate before any
rejection cuts are applied.

\subsection{Neutron Induced Background}\label{subsec:NIB}

Neutrons are a major source of background because their nuclear recoils render
them indistinguishable from WIMP events.  Nuclear recoils in the 10 keV range
arise from the elastic scattering of 0.1- 10 MeV neutrons on Xe. With an active
anti-coincidence, neutrons recoiling both in the LXe shield and the 
TPC LXe target are effectively identified and rejected. Nevertheless,
underground operation and a sufficient neutron shield to absorb or at least
thermalize the environment neutrons are essential for a WIMP experiment. The
main sources of neutrons are: 

\noindent {\em Muon Induced Neutrons} --
Cosmic ray muons produce neutrons in the Xe target and other materials by
spallation.  The primary concern is muon induced spallation of $^{136}$Xe and
$^{134}$Xe, leading to $^{135}$Xe and $^{133}$Xe respectively. These spallations
dominate because their abundances are about 10 times larger than for other Xe
isotopes. Both \g-rays and $\beta$ particles are produced in these
spallations.  We assume that the spallation
cross-section is $\sim 10$~mb.  With the neutron production rate for Homestake's
depth (4.4~kmwe), and accounting for the geometry of the detector, we calculate
a total neutron production rate of 0.01 cts/kg/d and a differential count rate
of $6\times10^{-5}$~cts/kg/d, before background rejection. This source of
background is therefore not a problem, and can be further reduced by a muon
veto. With a modest 99\% veto efficiency for muons, we expect to reduce the rate
to $<5\times10^{-6}$~cts/kg/d/keV. One is left then with neutrons from very high
energy muons, penetrating through the experiment. 

\noindent {\em $(\alpha,n)$ neutrons from surrounding rock} --
The incoming neutron flux, due to $(\alpha,n)$ reactions from U and Th decay in
the surrounding rock, will be $\sim 1000$ n/m$^2$/d.  To suppress this source of
background, a liquid scintillator shield is required. Typically, a 20~cm shield
can suppress the background to a level of $1\times10^{-5}$. The overall
background from this source is estimated at $\sim 1\times10^{-6}$~cts/kg/d/keV.

\noindent {\em Neutrons from U/Th contamination in the detector and surrounding
materials} --
Within the shielding, further neutrons arise from the U/Th in the materials of
the TPC  and of the shield and its vessel. 
A reasonable estimate based on current understanding of relevant materials is
$\sim 5\times10^{-5}$ cts/kg/d/keV and an ultimate goal, with higher purity
materials, would be $5\times10^{-6}$~cts/kg/d/keV. We choose the more aggressive
number for inclusion in the background  estimate.

\subsection{Gamma--Rays from PMTs}\label{subsec:GRPMT}

Gamma-ray background from the same U/Th and K in most detector materials will
dominate the overall count rate. PMTs, in particular, are typically a copious
source of gamma-rays. The K/U/Th content is highest in the HV divider chain. 
The baseline XENON LXeTPC uses special PMTs, stripped of the standard divider
chain and with selected compact metal envelopes and quartz windows. The
radioactivity is estimated at $\sim 100$ cts/d. 


\section{Experimental Sensitivity}\label{sec:sens}

\begin{figure}
\centering
\includegraphics[width=0.7\linewidth]{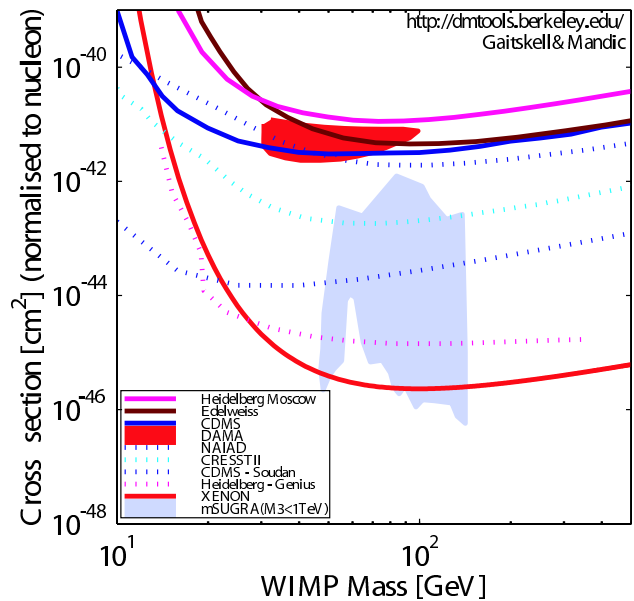}
\caption{Achieved and projected limits for spin independent couplings as a
function of WIMP mass \label{f:cross_section}} 
\end{figure}

The predicted sensitivity of the XENON experiment is shown in
Fig.~\ref{f:cross_section} along with that of other experiments: DAMA
\cite{DAMA00}, NAIAD \cite{spooner00}, HDMS\cite{HDMS99}, CDMS~I \cite{abusaidi00},
EDELWEISS \cite{benoit01}, CRESST~II \cite{bravin99}, CDMS~II
\cite{CDMS_Soudan}, GENIUS \cite{GENIUS98}. The prediction of minimal SUSY is
also plotted \cite{Mandic00}. The calculation has been done for a 1000~kg
fiducial target and for 3 years of operation. We have assumed that the
acceptance of the detector for real WIMPS is $\sim 1$. The assumed overall
background rate, as discussed in Sec.~\ref{sec:BKGD} is  $3.9 \times 10^{-5}$
cts/kg/d/keV. The fraction of misidentified candidate events was taken as
$\varepsilon_g = 0.5\%$, based on 99.5\% nuclear recoil discrimination in LXe.
  We have also assumed that the systematic
uncertainty on $\varepsilon_g$ is controlled at the 10\% level, and the
statistical limit therefore applies.  The energy band was taken as 10~\keV\ with
a visible energy threshold of 4~\keV.  We converted the visible energy to
recoil energy assuming that the quenching factor for xenon is 0.25. 
The optimal energy bandwidth depends on both the spectrum of the incident 
neutralinos as altered by the detector response matrix (including form factor
effects) and the detected background spectrum, which was assumed flat.


The sensitivity estimate presented in Fig.~\ref{f:cross_section} has large
potential sources of uncertainty which can act to either increase or decrease
it. The background estimates of Sec.~\ref{sec:BKGD} contain large
uncertainties in each of the leading terms. Detailed calculations and simulations will be 
undertaken to refine these estimates.  On the other
hand, our sensitivity estimate did not take into account the additional
background suppression expected from 3D event localization.
In addition, alternative photodetectors (LAAPD) or charge readout based on 
GEMs, also part our our studies, would have the advantage of substantially
lowering internal background or offsetting the sensitivity reducing terms
described above.  
An improvement of about a factor of 4 or more depending on readout 
scheme and the efficiency of the self-shield of the XENON detectors is
expected. 




\end{document}